\RequirePackage{lineno}
\documentclass[aps,prd,twocolumn,showpacs,superscriptaddress,showpacs]{revtex4}  
\usepackage{epsfig}
\usepackage{graphicx} 
\usepackage{dcolumn}  
\usepackage{bm}       
\usepackage{amssymb}  

\def\mevc {\ifmmode {\rm MeV}/c \else MeV$/c$\fi}
\def\mevcc {\ifmmode {\rm MeV}/c^2 \else MeV$/c^2$\fi}
\def\gevc {\ifmmode {\rm GeV}/c \else GeV$/c$\fi}
\def\gevcc {\ifmmode {\rm GeV}/c^2 \else GeV$/c^2$\fi}
\def\tevcc {\ifmmode {\rm TeV}/c^2 \else TeV$/c^2$\fi}
\def\ra   {\rightarrow}
\def\ol   {\overline}

\def\vtd  {\ifmmode |V_{td}| \else $|V_{td}|$\fi}
\def\vtb  {\ifmmode |V_{tb}| \else $|V_{tb}|$\fi}
\def\vts  {\ifmmode |V_{ts}| \else $|V_{ts}|$\fi}
\def\vcb  {\ifmmode |V_{cb}| \else $|V_{cb}|$\fi}

\newcommand{\Ds} {\ifmmode D_{\mbox{\sl s}}^{-}
                       \else $D_{\mbox{\sl s}}^{-}$\fi}
\newcommand{\Bs} {\ifmmode B_{\mbox{\sl s}}^{0}
                       \else $B_{\mbox{\sl s}}^{0}$\fi}
\newcommand{\Bsb} {\ifmmode \ol B_{\mbox{\sl s}}^{0}
                       \else $\ol B_{\mbox{\sl s}}^{0}$\fi}
\newcommand{\Bsh} {\ifmmode B_{\mbox{\sl s}}^H
                       \else $B_{\mbox{\sl s}}^H$\fi}
\newcommand{\Bsl} {\ifmmode B_{\mbox{\sl s}}^L
                       \else $B_{\mbox{\sl s}}^L$\fi}
\newcommand{\Dsl} {\ifmmode D_{\mbox{\sl s}}^{-} \ell^+
                       \else $D_{\mbox{\sl s}}^{-} \ell^+$\fi}
\newcommand{\xs} {\ifmmode x_{\mbox{\sl s}}
                       \else $x_{\mbox{\sl s}}$\fi}
\newcommand{\xd} {\ifmmode x_d \else $x_d$\fi}
\newcommand{\lxy} {\ifmmode L_{\rm xy} \else $L_{\rm xy}$\fi}
\newcommand{\dgam} {\ifmmode \Delta\Gamma \else $\Delta\Gamma$\fi}
\newcommand{\dm} {\ifmmode \Delta m \else $\Delta m$\fi}
\newcommand{\ctau} {\ifmmode c\tau \else $c\tau$\fi}
 
\newcommand{\et}{{ E}_{ T}}

\newcommand{\abseta}{|\eta| \leq}

\newcommand{\ptran}{\mbox{${p_T}$}}

\newcommand{\met}{\mbox{$\protect \raisebox{.3ex}{$\not$}\et$}}

\newcommand{\ppbar}{ { p} \bar{{ p}} }
\newcommand{\qqbar}{ { q} \bar{{ q}} }
\newcommand{\ttbar}{t\bar{t}}
\newcommand{\mttb}{M_{t\bar{t}}}

\newcommand{\mtop}{M_{t}}

\newcommand{\ipb}{ {\rm pb}^{-1} }

\begin{document}
\hspace{5.2in} \mbox{Fermilab-Pub-07-576-E}
\bibliographystyle{apsrev}


\title{Limits on the Production of Narrow $\ttbar$ Resonances
in $\ppbar$ Collisions at $\sqrt{s}=1.96$ TeV\\ }

\affiliation{Institute of Physics, Academia Sinica, Taipei, Taiwan 11529, Republic of China} 
\affiliation{Argonne National Laboratory, Argonne, Illinois 60439} 
\affiliation{Institut de Fisica d'Altes Energies, Universitat Autonoma de Barcelona, E-08193, Bellaterra (Barcelona), Spain} 
\affiliation{Baylor University, Waco, Texas  76798} 
\affiliation{Istituto Nazionale di Fisica Nucleare, University of Bologna, I-40127 Bologna, Italy} 
\affiliation{Brandeis University, Waltham, Massachusetts 02254} 
\affiliation{University of California, Davis, Davis, California  95616} 
\affiliation{University of California, Los Angeles, Los Angeles, California  90024} 
\affiliation{University of California, San Diego, La Jolla, California  92093} 
\affiliation{University of California, Santa Barbara, Santa Barbara, California 93106} 
\affiliation{Instituto de Fisica de Cantabria, CSIC-University of Cantabria, 39005 Santander, Spain} 
\affiliation{Carnegie Mellon University, Pittsburgh, PA  15213} 
\affiliation{Enrico Fermi Institute, University of Chicago, Chicago, Illinois 60637} 
\affiliation{Comenius University, 842 48 Bratislava, Slovakia; Institute of Experimental Physics, 040 01 Kosice, Slovakia} 
\affiliation{Joint Institute for Nuclear Research, RU-141980 Dubna, Russia} 
\affiliation{Duke University, Durham, North Carolina  27708} 
\affiliation{Fermi National Accelerator Laboratory, Batavia, Illinois 60510} 
\affiliation{University of Florida, Gainesville, Florida  32611} 
\affiliation{Laboratori Nazionali di Frascati, Istituto Nazionale di Fisica Nucleare, I-00044 Frascati, Italy} 
\affiliation{University of Geneva, CH-1211 Geneva 4, Switzerland} 
\affiliation{Glasgow University, Glasgow G12 8QQ, United Kingdom} 
\affiliation{Harvard University, Cambridge, Massachusetts 02138} 
\affiliation{Division of High Energy Physics, Department of Physics, University of Helsinki and Helsinki Institute of Physics, FIN-00014, Helsinki, Finland} 
\affiliation{University of Illinois, Urbana, Illinois 61801} 
\affiliation{The Johns Hopkins University, Baltimore, Maryland 21218} 
\affiliation{Institut f\"{u}r Experimentelle Kernphysik, Universit\"{a}t Karlsruhe, 76128 Karlsruhe, Germany} 
\affiliation{Center for High Energy Physics: Kyungpook National University, Daegu 702-701, Korea; Seoul National University, Seoul 151-742, Korea; Sungkyunkwan University, Suwon 440-746, Korea; Korea Institute of Science and Technology Information, Daejeon, 305-806, Korea; Chonnam National University, Gwangju, 500-757, Korea} 
\affiliation{Ernest Orlando Lawrence Berkeley National Laboratory, Berkeley, California 94720} 
\affiliation{University of Liverpool, Liverpool L69 7ZE, United Kingdom} 
\affiliation{University College London, London WC1E 6BT, United Kingdom} 
\affiliation{Centro de Investigaciones Energeticas Medioambientales y Tecnologicas, E-28040 Madrid, Spain} 
\affiliation{Massachusetts Institute of Technology, Cambridge, Massachusetts  02139} 
\affiliation{Institute of Particle Physics: McGill University, Montr\'{e}al, Canada H3A~2T8; and University of Toronto, Toronto, Canada M5S~1A7} 
\affiliation{University of Michigan, Ann Arbor, Michigan 48109} 
\affiliation{Michigan State University, East Lansing, Michigan  48824} 
\affiliation{University of New Mexico, Albuquerque, New Mexico 87131} 
\affiliation{Northwestern University, Evanston, Illinois  60208} 
\affiliation{The Ohio State University, Columbus, Ohio  43210} 
\affiliation{Okayama University, Okayama 700-8530, Japan} 
\affiliation{Osaka City University, Osaka 588, Japan} 
\affiliation{University of Oxford, Oxford OX1 3RH, United Kingdom} 
\affiliation{University of Padova, Istituto Nazionale di Fisica Nucleare, Sezione di Padova-Trento, I-35131 Padova, Italy} 
\affiliation{LPNHE, Universite Pierre et Marie Curie/IN2P3-CNRS, UMR7585, Paris, F-75252 France} 
\affiliation{University of Pennsylvania, Philadelphia, Pennsylvania 19104} 
\affiliation{Istituto Nazionale di Fisica Nucleare Pisa, Universities of Pisa, Siena and Scuola Normale Superiore, I-56127 Pisa, Italy} 
\affiliation{University of Pittsburgh, Pittsburgh, Pennsylvania 15260} 
\affiliation{Purdue University, West Lafayette, Indiana 47907} 
\affiliation{University of Rochester, Rochester, New York 14627} 
\affiliation{The Rockefeller University, New York, New York 10021} 
\affiliation{Istituto Nazionale di Fisica Nucleare, Sezione di Roma 1, University of Rome ``La Sapienza," I-00185 Roma, Italy} 
\affiliation{Rutgers University, Piscataway, New Jersey 08855} 
\affiliation{Texas A\&M University, College Station, Texas 77843} 
\affiliation{Istituto Nazionale di Fisica Nucleare, University of Trieste/\ Udine, Italy} 
\affiliation{University of Tsukuba, Tsukuba, Ibaraki 305, Japan} 
\affiliation{Tufts University, Medford, Massachusetts 02155} 
\affiliation{Waseda University, Tokyo 169, Japan} 
\affiliation{Wayne State University, Detroit, Michigan  48201} 
\affiliation{University of Wisconsin, Madison, Wisconsin 53706} 
\affiliation{Yale University, New Haven, Connecticut 06520} 
\author{T.~Aaltonen}
\affiliation{Division of High Energy Physics, Department of Physics, University of Helsinki and Helsinki Institute of Physics, FIN-00014, Helsinki, Finland}
\author{J.~Adelman}
\affiliation{Enrico Fermi Institute, University of Chicago, Chicago, Illinois 60637}
\author{T.~Akimoto}
\affiliation{University of Tsukuba, Tsukuba, Ibaraki 305, Japan}
\author{M.G.~Albrow}
\affiliation{Fermi National Accelerator Laboratory, Batavia, Illinois 60510}
\author{B.~\'{A}lvarez~Gonz\'{a}lez}
\affiliation{Instituto de Fisica de Cantabria, CSIC-University of Cantabria, 39005 Santander, Spain}
\author{S.~Amerio}
\affiliation{University of Padova, Istituto Nazionale di Fisica Nucleare, Sezione di Padova-Trento, I-35131 Padova, Italy}
\author{D.~Amidei}
\affiliation{University of Michigan, Ann Arbor, Michigan 48109}
\author{A.~Anastassov}
\affiliation{Rutgers University, Piscataway, New Jersey 08855}
\author{A.~Annovi}
\affiliation{Laboratori Nazionali di Frascati, Istituto Nazionale di Fisica Nucleare, I-00044 Frascati, Italy}
\author{J.~Antos}
\affiliation{Comenius University, 842 48 Bratislava, Slovakia; Institute of Experimental Physics, 040 01 Kosice, Slovakia}
% ...
\collaboration{CDF Collaboration\footnote{With visitors from $^a$University of Athens, 15784 Athens, Greece, 
$^b$Chinese Academy of Sciences, Beijing 100864, China, 
$^c$University of Bristol, Bristol BS8 1TL, United Kingdom, 
$^d$University Libre de Bruxelles, B-1050 Brussels, Belgium, 
$^e$University of California Irvine, Irvine, CA  92697, 
$^f$University of California Santa Cruz, Santa Cruz, CA  95064, 
$^g$Cornell University, Ithaca, NY  14853, 
$^h$University of Cyprus, Nicosia CY-1678, Cyprus, 
$^i$University College Dublin, Dublin 4, Ireland, 
$^j$University of Edinburgh, Edinburgh EH9 3JZ, United Kingdom, 
$^k$University of Heidelberg, D-69120 Heidelberg, Germany, 
$^l$Universidad Iberoamericana, Mexico D.F., Mexico, 
$^m$University of Manchester, Manchester M13 9PL, England, 
$^n$Nagasaki Institute of Applied Science, Nagasaki, Japan, 
$^o$University de Oviedo, E-33007 Oviedo, Spain, 
$^p$Queen Mary, University of London, London, E1 4NS, England, 
$^q$Texas Tech University, Lubbock, TX  79409, 
$^r$IFIC(CSIC-Universitat de Valencia), 46071 Valencia, Spain, 
}}
\noaffiliation

\affiliation{URL: http://www-cdf.fnal.gov}

\begin{abstract}
We search for evidence of resonant top quark pair production in 955 $\ipb$ of $\ppbar$ collisions at $\sqrt{s}=1.96$ TeV recorded with the CDF II detector at the Fermilab Tevatron. For fully reconstructed candidate $\ttbar$ events triggered on leptons with large transverse momentum and containing at least one identified $b$-quark jet, we compare the invariant mass spectrum of $\ttbar$ pairs to the expected superposition of standard model $\ttbar$, non-$\ttbar$ backgrounds, and a simple resonance model based on a sequential $Z'$ boson. We establish upper limits for $\sigma(\ppbar\ra Z')\cdot Br(Z' \rightarrow \ttbar)$ in the $Z'$ mass interval from 450 $\gevcc$ to 900 $\gevcc$. A topcolor leptophobic $Z'$ is ruled out below 720 $\gevcc$, and the cross section of any narrow $Z'$-like state decaying to $\ttbar$ is found to be less than 0.64 pb at 95\% C.L. for $M_{Z'}$ above 700 $\gevcc$.

\end{abstract}


\maketitle

\date{\today}

Resonant top pair production in hadronic collisions has been discussed in the context of extended gauge theories with massive $Z$-like bosons \cite{r1,r2,r3}, in theories with topcolor \cite{r4}, or with axigluons \cite{r5}. Decays to $\ttbar$ are of special interest in leptophobic models that would evade detection in traditional searches based on dielectron or dimuon signatures. More recently, resonant top pairs have been suggested as signatures for Kaluza-Klein (KK) states of gluons, weak bosons, and gravitons \cite{kkg,kkw,kkgr}; in some of these models the KK excitation couples strongly to the top quark and $\ttbar$ is the dominant decay mode.

A $\ttbar$ resonance would appear as unexpected structure in the spectrum of the invariant mass of $\ttbar$ 
pairs $M_{\ttbar}$. Previous searches using  $\approx 100 ~\ipb$ samples from Fermilab Tevatron Run I have ruled out the production 
of a narrow leptophobic topcolor resonance with mass less than 480~$\gevcc$ ~\cite{runIcdf,runId0}. Here, we search for resonant structure in the $M_{\ttbar}$ spectrum in 955 $\ipb$ of $\ppbar$ collisions at {$\sqrt{s}=1.96$} TeV recorded with the CDF II detector in Tevatron Run II. Modeling the resonance as a narrow massive vector boson $Z'$, and calculating its mass with techniques used in precision measurement of the top quark mass \cite{topmass}, we set limits on the cross section times branching ratio $\sigma B=\sigma(\ppbar\ra Z')\cdot Br(Z'\ra\ttbar)$ as a function of $M_{Z'}$. This study is complementary to Ref. \cite{flame}, which uses a different event selection and reconstruction of the $\ttbar$ kinematics. 

The CDF II detector comprises a spectrometer in a 1.4 T magnetic field surrounded by projective electromagnetic and hadronic calorimeters and muon detectors \cite{CDF}. The spectrometer, consisting of silicon microstrip detectors surrounded by a large open cell drift chamber, provides precision track reconstruction and displaced secondary vertex detection. We use coordinates where $\phi$ is the azimuthal angle, $\theta$ is the polar angle with respect to the proton beam axis, transverse energy is $E_T = E\sin(\theta)$, and the pseudorapidity is $\rm{\eta = -ln[tan(\theta/2)]}$. The data used here were recorded between March 2002 and January 2006.  

We collect a sample of $\ttbar \rightarrow W^{+}bW^{-}\bar{b}$ candidate events with one leptonic $W$ boson decay using triggers that require a central ($\abseta 1.0$) electron with $E_T>$18~GeV or central muon with transverse momentum $p_T>$18~GeV/$c$. After offline reconstruction, we select events with an isolated electron with $E_T\geq 20$~GeV or muon with $p_T\geq 20$~GeV/c, missing transverse energy $\met \geq~$20~GeV consistent with a neutrino from $W$ decay, and at least four hadronic jets with $\abseta$~2.0, of which three must have $\et\geq 15$~GeV, and a fourth must have $\et \geq 8$~GeV \cite{topxsec}. The jets are clustered in fixed cones of radius $\Delta R = \sqrt{(\Delta \eta)^2 + (\Delta \phi)^2} \leq 0.4$. At least one of the jets is required to be $b$-tagged, i.e. contain a reconstructed secondary vertex displaced from the primary event vertex as expected from the decay of a bottom hadron in the jet \cite {g4topxsec}. We find 347 events fulfilling these criteria. 

The sample is dominated by $s$-channel $\qqbar$ annihilation into $\ttbar$ pairs ~\cite{topxsectheory1, topxsectheory2}. The $\ttbar$ acceptance and efficiencies are calculated using the {\sc herwig} generator~\cite{herwig} and a detector simulation, assuming a top mass $\mtop = 175 ~\gevcc$. The simulated detector response, particularly with respect to lepton isolation, jet energies, and $b$-tagging, has been tuned in an earlier measurement of the top pair production cross section \cite{topxsec}. The total combined trigger and reconstruction efficiency is $3.5\pm 0.5\%$.  Non-$\ttbar$ backgrounds include $W$ bosons produced in association with jets ($W$+jets), where a light flavor jet is incorrectly $b$-tagged; $W$+jets events with real heavy-flavor jets; mismeasured QCD multi-jet events with one jet identified as a lepton; and smaller contributions from electroweak processes such as diboson ($WW, WZ, ZZ$) and single--top production. The rates and kinematics of these processes are modeled with simulated and data control samples as employed in the top cross section measurement \cite{topxsec}.  A total of $73 \pm 9$ non-$\ttbar$ background events are expected.

The final state of four jets, a high-$\ptran$ lepton, and $\met$ allows an over-constrained reconstruction of the top pair kinematics. The assignment of jets to quarks most consistent with the $\ttbar$  hypothesis is determined using the $\chi^2$ minimization algorithm employed in the  measurement of the top mass \cite{topmass}. Here, following \cite{liss}, we include the known top mass as a constraint, which improves the accuracy of the parton assignments. The measured jet energies are corrected back to parton values using calibrations derived from photon-jet balancing and detector simulation \cite{jes}. In the $\chi^2$ minimization the parton energies are varied within their uncertainties and the $W$ and top masses are constrained to the values $M_W = 80.4 ~\gevcc$ and $M_t = 175.0 ~\gevcc$ within their natural widths (2.1 and 1.5 $~\gevcc$ respectively). The effect of variation in the central value of $M_t$ is included later as a systematic uncertainty. Jets with $b$-tags must be associated with $b$ quarks.  The jet-quark assignment giving the lowest $\chi^2$ consistent with these constraints is chosen as the solution. In simulated $\ttbar$ events we find a small number of poorly reconstructed events flagged by extreme $\chi^2$. We find the sensitivity of the search is optimized by requiring $\chi^2 < 50$; this cut removes 4\% of $\ttbar$ events and 9\% of non-$\ttbar$ backgrounds.

We model the resonant $\ttbar$ production mechanism as a sequential $Z'$, a heavy neutral boson with the same couplings as the $Z$, here including decay to $\ttbar$ with $M_{t}= 175 ~\gevcc$. This electroweak channel has no interference with the strong $\ttbar$ production processes and the model lineshape is purely Lorentzian.  To facilitate comparison to other results \cite{flame,runIcdf,runId0} we assign the same narrow width used there, $\Gamma_{Z'}=0.012M_{Z'}$. A strictly sequential $Z'$ with open $\ttbar$ decays has  $\Gamma_{Z'}\simeq 0.03M_{Z'}$. Since our reconstructed mass resolution is greater than 60~$\gevcc$ (see below) the analysis is insensitive to model dependent width differences at this level, and applies to any narrow $\ttbar$ state appearing as a single enhancement in the $M_{\ttbar}$ spectrum. Signal models are generated using the {\sc pythia} simulation \cite{pythia} with $Z'$ masses between 450 and 900 $\gevcc$ in increments of 50 \gevcc.

\begin{figure}[ht]
\begin{center}
\includegraphics[width=1.0\columnwidth]{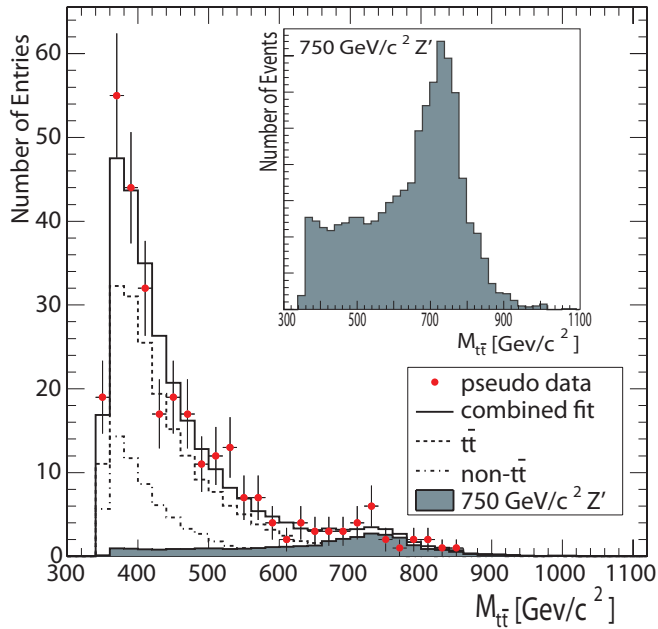}
\caption{Simulated $M_{\ttbar}$ spectrum for 955~$\ipb$ in presence of a 750 $\gevcc ~Z'$ with $\sigma B= 1$~pb (shaded curve). The points are a simulated data set. The solid line is the best fit to a superposition of the $Z'$ signal (solid histogram) and the expected $\ttbar$ (dot) and non-$\ttbar$ (dot-dash) backgrounds. The inset shows a magnified view of the reconstructed $M_{\ttbar}$ spectrum for a 750 $\gevcc$ $Z'$. The low mass tail arises from incorrect jet-parton associations. \label{Expected}}
\end{center}
\end{figure}
The inset of Fig.~\ref{Expected} shows the $M_{\ttbar}$ distribution reconstructed for a simulated 750 $\gevcc$ $Z'$. There is a peak near the expected value and a low mass tail which arises from the incorrect jet-parton assignments where a jet from initial or final state radiation has been used instead of a jet from top decay. The rms of the peak region is approximately 60 $\gevcc$ and the full rms is 137 $\gevcc$. Other $Z'$ masses show similar behavior: the $M_{Z'}$ peak width is preserved and the low mass tail extends down to the kinematic threshold at 350 $\gevcc$. The full rms of the $M_{\ttbar}$ distribution varies between 67 and 178 $\gevcc$ over our $Z'$ mass range. The fraction of $Z'$ removed by the $\chi^2$ cut varies between 4\% and 9\% over the $Z'$ mass range.
   
We use a three-parameter binned likelihood maximization to fit the $M_{t\bar{t}}$ ~spectrum to a superposition of the expected shapes for ${Z'} \rightarrow t\bar{t}$, standard model $t\bar{t}$, and non-$t\bar{t}$ processes. In the ${i}^{\rm th}$ bin, we expect 

\begin{equation}
\mu_{i}=\left[{\sigma B}{A}{\epsilon}\int{\mathcal{L}}{dt} \right]{ {P}_{Z',i}} + {N}_{t\bar{t}}{ {P}_{t\bar{t},i}} + {N}_{\rm bkg}{ {P}_{{\rm bkg},i}}
\label{MuI}
\end{equation}
\noindent
where ${P}_{Z',i}$, ${P}_{t\bar{t},i}$, and ${P}_{{\rm bkg},i}$ are the probabilities of observing a signal event, $t\bar{t}$ event or non-$t\bar{t}$ background event in bin $i$, respectively. ${N}_{t\bar{t}}$ and ${N}_{\rm bkg}$ are the number of non-resonant $t\bar{t}$ and the non-$t\bar{t}$ background events. The ${\sigma B}{A}{\epsilon}\int{\mathcal{L}}{dt}$ term contains the product of cross-section and $\ttbar$ branching ratio, acceptance, and efficiency for the $Z'$, and the luminosity. 

A likelihood function ${L}$ for the distribution can be written as

\begin{center}
\begin{equation}
{{L}}=\prod_{{i,k}}\mathcal{P}_{i}({{n}_{i}}|{\mu_{i}}) {G}({\nu_{ k}}|{\bar{\nu}_{ k},\sigma_{\nu_{ k}} }).
\label{LkFcn}
\end{equation}
\end{center}
\noindent
The function $\mathcal{P}_{i}({{n}_{i}}|{\mu_{i}})$ is the Poisson probability for observing $n_i$ events in a bin $i$ where $\mu_i $ are expected. The functions ${G}({\nu_{ k}}|{\bar{\nu}_{ k},\sigma_{\nu_{ k}} })$ constrain the nuisance parameters $\nu_{k}$, which include the non-$\ttbar$ background normalization $N_{\rm bkg}$, $b$-tag efficiency, acceptances and luminosities, with Gaussian probability around their central values ${\bar{\nu}_{ k}}$ and uncertainties $\sigma_{\nu_{ k}}$. The $\ttbar$ and non-$\ttbar$ background values are taken from \cite{topxsec}, and the $Z'$ acceptances and efficiencies are determined from the {\sc pythia} simulation. We find ${\sigma B}$, ${N}_{t\bar{t}}$, ${N}_{\rm bkg}$, and ${\nu_{ k}}$ that maximize the likelihood function for each $M_{Z'}$.

The algorithm is tested with simulated samples where the $\ttbar$, non-$\ttbar$, and $Z'$ models are combined in the expected ratios and sampled with the expected level of statistical fluctuations. The points in the main part of Fig.~\ref{Expected} show the ${M_{t\bar{t}}}$ distribution for a
simulated data sample corresponding to an integrated luminosity of 955~$\ipb$ in the case of a 750 $\gevcc$ $Z'$ with $\sigma B$ = 1 pb. The histograms show the components as resolved by the likelihood fit. 
The extraction of the $Z'$ component uses shape information from the low mass part of the specturm as well as the peak area. 

The 95\% C.L. upper limit on $\sigma B$ at a given mass is found by integrating the likelihood along ${\sigma B}$, re-optimizing at each point, to find the  value that contains 95\% of the area.  We measure our expected sensitivity using large ensembles of simulated samples like the one shown in Fig.~\ref{Expected}. The main sources of systematic uncertainty are the acceptance change due to energy scale uncertainty on the jet thresholds, and the shape change in $M_{\ttbar}$ from the top mass uncertainty. Model dependent shape effects associated with initial and final state gluon radiation and non-$\ttbar$ backgrounds are small. Simulated samples with reasonable variations for systematic effects are used to measure the apparent shifts in the fitted $\sigma B$ as a function of the true value. The sum of the shifts in quadrature is used as the width of a Gaussian resolution function that is convolved with the likelihood as a function of $\sigma B$. The systematic uncertainties worsen the limits by roughly 0.2 pb, independent of 
the $Z'$ mass, with the increase dominated by the effects of jet energy scale and the top mass uncertainty in equal measure. The expected 95\% C.L. upper limits including all sources of uncertainty are shown as a function of $M_{Z'}$ in the middle column of Table~\ref{LimitTable}. If no $Z'$ is present our expected cross
section limit at high $M_{Z'}$ is 0.55 pb.  

\begin{figure}[ht]
\begin{center}
\includegraphics[width=1.0\columnwidth]{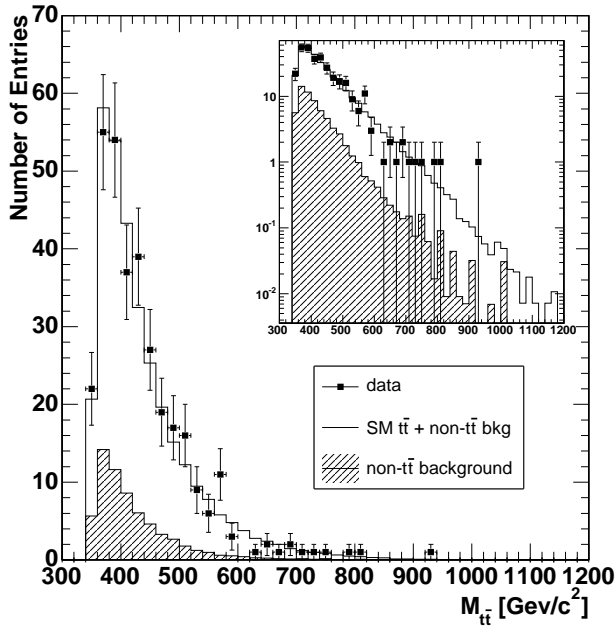}
\caption{The invariant mass of top quark pairs $M_{\ttbar}$ observed in the data is compared to the no $Z'$ expectation. The non-$\ttbar$ backgrounds are constrained to the expected value and the sum of $\ttbar$ and non-$\ttbar$ equal the number of data events.\label{mttblin}}
\end{center}
\end{figure}

The $\mttb$ distribution measured in the data is shown in Fig.~\ref{mttblin}. A final sample of 327 candidates remains after the $\chi^2$ requirement. In this figure we compare the observation to the expected spectrum in the case of no $Z'$. The non-$\ttbar$ component is fixed at the expected value and the $\ttbar$ normalization is scaled to match the total number of events. The inferred top production cross section is $\sigma(\ttbar) = 7.8 \pm 0.7$ pb (statistical error only), to be compared with the predicted standard model value of 6.7~pb for $M_{t}= 175 ~\gevcc$ \cite{topxsectheory1, topxsectheory2}. The inset shows the measurement on a logarithmic scale. The simulated $M_{\ttbar}$ spectra for $\ttbar$ and non-$\ttbar$ describe the data well. 
 
Applying the full limit procedure to the spectrum  
in Fig.~\ref{mttblin} we find 95\% C.L. upper limits on $\sigma(\ppbar\ra Z')\cdot Br(Z' \rightarrow \ttbar)$ as listed in the rightmost column of Table \ref{LimitTable}. The limits at high mass are consistent with expectation. At lower masses our measurement shows an excursion above the expected value of approximately one standard deviation.
\begin{table}[ht]
\caption{Expected and observed limits (95\% C.L.) on ${\sigma(\ppbar\ra Z')\cdot Br(Z'\rightarrow \ttbar)}$ as a function of ${M_{Z'}}$ for 955 $\ipb$, including both statistical and systematic uncertainties.}
\begin{ruledtabular}
\begin{tabular}{ccc}
${M_{Z'} (\gevcc)}$ & Expected Limit (pb) & Observed Limit (pb)\\
\hline
\vspace*{-0.3cm}\\
450 & $2.27_{-0.57}^{+0.79}$ & 3.39 \\
500 & $1.92_{-0.40}^{+0.63} $ & 2.72\\

550 & $1.37_{-0.30}^{+0.45}$ & 1.57\\
600 & $0.97_{-0.18}^{+0.33}$ & 0.83\\
650 & $0.78_{-0.13}^{+0.24}$ & 0.65\\
700 & $0.70_{-0.12}^{+0.14}$  & 0.64\\
750 & $0.64_{-0.11}^{+0.15}$ & 0.61\\
800 & $0.58_{-0.07}^{+0.15}$ & 0.60\\
850 & $0.55_{-0.05}^{+0.10}$  & 0.57\\
900 & $0.55_{-0.06}^{+0.08} $  & 0.57\\
\end{tabular}
\end{ruledtabular}
\label{LimitTable} \rm
\end{table}

\begin{figure}[ht]
\begin{center}
\includegraphics[width=1.0\columnwidth]{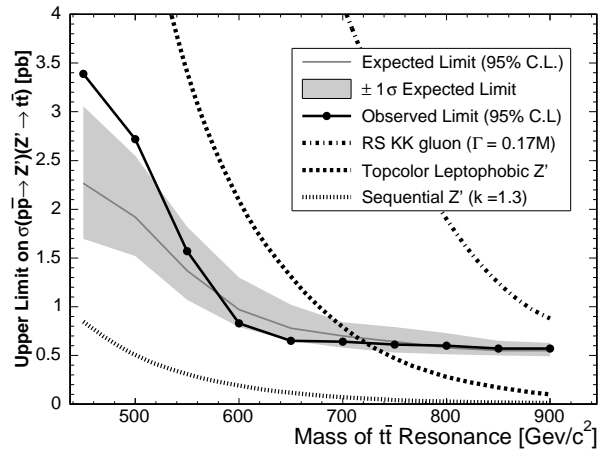}
\caption{Upper limits (95\% C.L.) on the production cross section for $\ttbar$ resonance along with expected cross sections for several models.\label{Data_Limit}}
\end{center}
\end{figure}

The result is represented graphically and compared to some theoretical predictions in Fig.~\ref{Data_Limit}. The observed limit is the solid black line and the shaded band around the grey line denotes the $\pm 1\sigma$ uncertainties around the expected upper limit. A leptophobic $Z'$ predicted by the topcolor theory \cite{r4}, shown as a large-dotted line, is ruled out below 720 $\gevcc$ at 95\% C.L. The small-dotted curve at the bottom of the figure is the expected cross section for a sequential $Z'$, calculated with the {\sc herwig} simulation using a multiplicative factor of 1.3 to account for NLO effects. A leptophobic $Z'$ with these couplings would evade direct searches in dilepton final states, and because the $\ttbar$ detection efficiency is small, is still out of range of our sensitivity in the $\ttbar$ mode. The Tevatron cross section for the KK gluon excitation in the Randall-Sundrum model of Ref.~\cite{kkg} is shown as a dot-dash line \cite{ben}. Since the KK resonance is broad ($\Gamma \approx 0.17M$), our limits derived in the ``narrow width'' assumption are not strictly applicable; we show the curve here for qualitative comparison. The cross section of any narrow $Z'$-like state produced in $\ppbar$ collisions at $\sqrt{s}=1.96$ TeV and subsequently decaying to $\ttbar$ is less than or equal to 0.64~pb (95\% C.L) for all $M_{Z'}$ above 600 \gevcc. 

\begin{acknowledgments}
We thank the Fermilab staff and the technical staffs of the participating institutions for their vital contributions. This work was supported by the U.S. Department of Energy and National Science Foundation; the Italian Istituto Nazionale di Fisica Nucleare; the Ministry of Education, Culture, Sports, Science and Technology of Japan; the Natural Sciences and Engineering Research Council of Canada; the National Science Council of the Republic of China; the Swiss National Science Foundation; the A.P. Sloan Foundation; the Bundesministerium f\"ur Bildung und Forschung, Germany; the Korean Science and Engineering Foundation and the Korean Research Foundation; the Science and Technology Facilities Council and the Royal Society, UK; the Institut National de Physique Nucleaire et Physique des Particules/CNRS; the Russian Foundation for Basic Research; the Comisi\'on Interministerial de Ciencia y Tecnolog\'{\i}a, Spain; the European Community's Human Potential Programme; the Slovak R\&D Agency; and the Academy of Finland.
\end{acknowledgments}

\newpage

\bibliography{prd}

\end{document}